\documentclass[aps, onecolumn, superscriptaddress]{revtex4}
\usepackage{cmap}
\usepackage{mdwlist}

\usepackage[version=3]{mhchem} % Formula subscripts using \ce{}
\usepackage{graphicx}
\usepackage{amsmath,amssymb}

\input{epsf}
\input{epsfx}
\def\bra#1{\mathinner{\langle{#1}|}}
\def\ket#1{\mathinner{|{#1}\rangle}}
\def\braket#1{\mathinner{\langle{#1}\rangle}}

\newcommand{\fr}[2]{\frac{{#1}}{{#2}}}

\newcommand{\unit}[1]{\hspace{-1.3pt} {#1}}

\begin{document}

\title{Deterministic coupling of a single nitrogen vacancy center to a photonic crystal cavity\\}
%
%%
%\author{Dirk Englund$^{1,*,\dagger}$, Brendan Shields$^{1,*}$, Kelley Rivoire$^{2}$,  Fariba Hatami$^{3}$,  Jelena Vu\v{c}kovi\'{c}$^{2}$, Hongkun Park$^{1,4,\dagger}$, Mikhail D. Lukin$^{1,\dagger}$}

%\maketitle
%\begin{itemize*}
% \item[1. ] Department of Physics, Harvard University, Cambridge MA 02138
% \item[2. ] Department of Electrical Engineering, Stanford University, Stanford CA 94305
% \item[3. ] Department of Physics, Humboldt-Universit\"at zu Berlin, Newtonstra\ss{}e 15, 12489 Berlin%, Germany
% \item[4. ] Department of Chemistry and Chemical Biology, Harvard University, Cambridge MA 02138

%\item[*] These authors contributed equally.

%\end{itemize*}

%
%%

%\altaffiliation{Current address: Department of Electrical Engineering, Columbia University, New York, NY 10027 }

\author{Dirk Englund\footnote{Current address: Columbia University, New York, NY. Email: englund$@$columbia.edu}}
\affiliation{Department of Physics, Harvard University, Cambridge MA 02138}

\affiliation{These authors contributed equally.}
\author{Brendan Shields}
\affiliation{Department of Physics, Harvard University, Cambridge MA 02138}
\affiliation{These authors contributed equally.}
\author{Kelley Rivoire}
\affiliation{Department of Electrical Engineering, Stanford University, Stanford CA 94305 }
\author{Fariba Hatami}
\affiliation{Department of Physics, Humboldt-Universit\"at zu Berlin, Newtonstra\ss{}e 15, 12489 Berlin }
\author{Jelena Vu\v{c}kovi\'{c}}
\affiliation{Department of Electrical Engineering, Stanford University, Stanford CA 94305 }
\author{Hongkun Park}\email{Hongkun_Park@harvard.edu}
\affiliation{Department of Physics, Harvard University, Cambridge MA 02138}
\affiliation{Department of Chemistry and Chemical Biology, Harvard University, Cambridge MA 02138}
\author{Mikhail D. Lukin}\email{lukin@physics.harvard.edu}
\affiliation{Department of Physics, Harvard University, Cambridge MA 02138}

\begin{abstract}

We describe and experimentally demonstrate a technique for deterministic coupling between a photonic crystal (PC) nanocavity and single emitters. The technique is based on in-situ scanning of a PC cavity over a sample and allows the positioning of the cavity over a desired emitter with nanoscale resolution. The power of the technique, which we term a Scanning Cavity Microscope (SCM), is demonstrated by coupling the PC nanocavity to a single nitrogen vacancy (NV) center in diamond, an emitter system that provides optically accessible electron and nuclear spin qubits.

%We describe and experimentally demonstrate a technique for deterministic, large coupling between photonic crystal (PC) nanocavity and single photon emitters. The technique is based on in-situ scanning  of PC cavity over a sample that allows to position the cavity over the desired emitter with nanoscale resolution. The technique is applied to couple the PC nanocavity to a single nitrogen vacancy (NV) center in diamond, an emitter system that provides optically accessible electron and nuclear spin qubits that are promising for applications in quantum information science.

%We describe an in-situ method for scanning a photonic crystal nanocavity over a sample, providing a positionable quantum optical interface to study and control the radiative properties of solid state emitters. We apply this technique to deterministically couple the nanocavity to a single nitrogen vacancy (NV) center in diamond, an emitter system that provides optically accessible electron and nuclear spins that are promising as solid state qubits. 

\end{abstract}

\maketitle

Optical resonators enable large amplification of small optical signals, resulting in a range of spectroscopic and sensing applications, and have allowed for detection of single atoms\cite{2000.Science.Hood_Kimble.Atom_cavity_microscope}, molecules \cite{2007.Science.Armani-Vahala.sensing}, and quantum dots \cite{2007.Nature1,2007.Nature.Srinivasan-Painter}. In addition, they enable a controllable coupling between optical emitters and the cavity vacuum field that is critical for efficient light sources \cite{2005.Science.Noda.SE_control,2005.PRL.Englund,2007.NPhoton.Strauf} and for the realization of memory nodes in quantum networks \cite{CZKM1997PRL} and quantum repeaters \cite{1998.PRL.Briegel.quantum-repeater}. This coupling strength scales with the cavity mode volume $V_{m}$ as $1/\sqrt{V_{m}}$, and consequently, nanoscale photonic crystal (PC) cavities have been explored extensively in solid-state cavity QED applications. While much progress has been achieved in coupling quantum dots to PC cavities made from the host material \cite{2004.Nature.Yoshie,2007.Nature.Hennessy-Imamoglu.Strong_coupling_quantum_nature,2007.Nature1}, extending these techniques to fully deterministic coupling and to other material systems has been difficult. Specifically, there has been much recent interest in coupling PC resonators to NV centers \cite{2009.OptLett.Benson.diamond_NC_coupling_PC,2009.OpEx.Barclay-Beasoleil.PC_NV_hybrid,2008.OpEx.Bayn_Salzman.NV_diamond_cavity,2008.OpEx.Kreuzer_Becher.diamond_cav}, a promising single photon emitter with excellent electronic and nuclear spin memory \cite{2008.Science.Neumann-Wrachtrup.spin_entanglement,2009.Science.Jiang-Lukin.repetitive_readout_spin_NV,2009.Science.Awschalom.NV_spin_GHz_dynamics}, though experimental demonstrations have remained a challenge.

%Optical resonators enable large amplification of small optical signals, resulting in a range of spectroscopic and sensing applications and has allowed for detection of single atoms\cite{2000.Science.Hood_Kimble.Atom_cavity_microscope,1992.PRL.Thompson-Kimble.Normal_mode_splitting}, molecules\cite{2007.Science.Armani-Vahala.sensing}, and quantum dots\cite{2007.Nature1,2007.Nature.Srinivasan-Painter}. In addition, they enable a strong coupling between optical emitters and the cavity vacuum field that are critical for generating efficient light sources\cite{2005.Science.Noda.SE_control,2005.PRL.Englund,2007.NPhoton.Strauf} and quantum information processing modules\cite{CZKM1997PRL}. This coupling strength scales with the cavity mode volume $V_{m}$ as $1/\sqrt{V_{m}}$, and consequently, photonic crystal (PC) nanocavities have been explored extensively in solid-state cavity QED applications. Specifically, there has been much recent interest in coupling PC resonators to NV centers\cite{2009.OptLett.Benson.diamond_NC_coupling_PC,2009.OpEx.Barclay-Beasoleil.PC_NV_hybrid,2008.OpEx.Bayn_Salzman.NV_diamond_cavity,2008.OpEx.Kreuzer_Becher.diamond_cav,2008.OpEx.Loncar.NV_PC_design}, although experimental demonstrations have remained a challenge. 

\begin{figure}
\centering{\includegraphics[width=\linewidth]{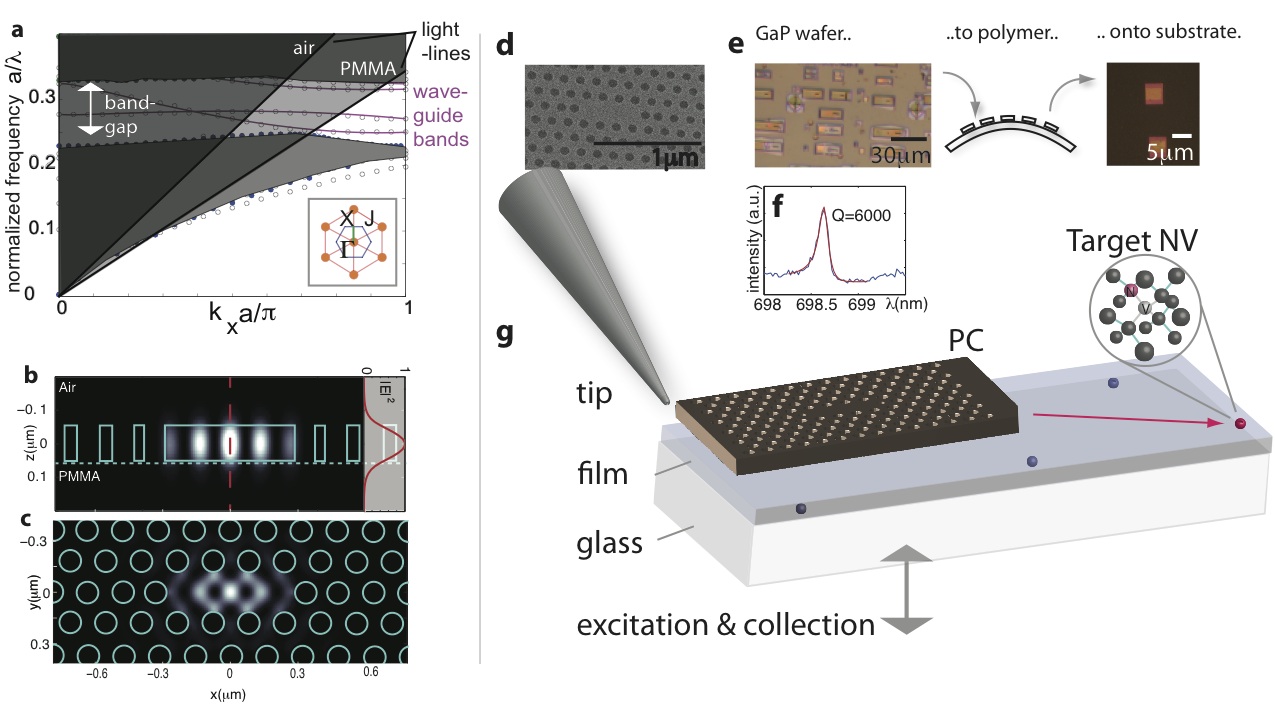}}
\caption{{\footnotesize  (a) Dispersion for the photonic crystal slab in air, along the waveguide direction $k_x$; the inset shows the crystal (orange) and inverse crystal (blue) directions. The lattice has a periodicity of $a=176$\unit{nm}, hole radius of $\sim 53$\unit{nm}, and slab height of $\sim 110$\unit{nm}.  (b) Energy density for fundamental mode in cross-section and (c) in plane. (d) SEM. (e) The photonic crystals are transferred from the GaP chip onto a substrate via a polymer stamp. (f) Broad-band reflectivity measurement of a cavity resonance with $Q\approx 6\cdot 10^{3}$. (g) The PC slab is positioned relative to a target nanocrystal in the polymer film.}}\label{fig:setup}
%\vskip2mm
\end{figure}

We demonstrate a technique for deterministic positioning  of micron-scale PC slabs that support high quality factor ($Q$) cavity modes with nanometer-scale features. When such a cavity is scanned over the sample, it can be used for deterministically coupling to optically active systems with sub-wavelength resolution via the evanescent field.  By appropriate design of PC cavities and waveguides, these systems combine sub-wavelength resolution, high throughput, and cavity-enhanced sensitivity. In particular, they can be deterministically interfaced with isolated optical emitters. 

In our experiments, the PC consists of a triangular lattice of air holes in a gallium phosphide (GaP) membrane, creating an optical bandgap that confines light in the slab to a cavity region. The bandgap along the $\Gamma X$ crystal direction is shown in the dispersion diagram in Figure \ref{fig:setup}(a). Confinement in the vertical direction occurs through total internal reflection (TIR) for modes with frequencies below the air light-line indicated in Figure \ref{fig:setup}(a). A row of missing holes supports band modes which form bound cavity states when terminated on two sides.  We employ a three-hole defect cavity \cite{Noda2003Nature} whose geometry is optimized for use on a Poly(methyl methacrylate) (PMMA) substrate with a refractive index of $n_{s}\sim 1.5$. The TIR-confined region in k-space is smaller on top of the PMMA, as sketched in Figure \ref{fig:setup}(a), but simulations indicate that the $Q$ value can still be above $13\cdot 10^{3}$. The cavity has a mode volume $V_{m}=0.74 (\lambda/n_{\mbox{\tiny{GaP}}})^{3}$, where $n_{\mbox{\tiny{GaP}}}=3.4$ is the refractive index of GaP at $\lambda=670$\unit{nm}.  The fundamental mode of the PC cavity is depicted by its energy density in Figure \ref{fig:setup}(c). The cross section in Figure \ref{fig:setup}(b) shows the evanescent tail of the mode that couples to emitters. 

We fabricate GaP PC nanocavities by a combination of electron beam lithography and dry etching \cite{2008.APL.Rivoire-Vuckovic.GaP_PC} of a 108\unit{nm} membrane of GaP on top of a 940\unit{nm}-thick sacrificial layer of a Al$_{0.85}$Ga$_{0.15}$P. A wet etch removes the sacrificial layer, leaving free-standing photonic crystal membranes. The scanning electron micrograph (SEM) of a resulting PC nanocavity is shown in Figure \ref{fig:setup}(d). Reflectivity measurements of freestanding cavities indicate that quality factors (Q) of these cavities can exceed $6\cdot 10^{3}$, the maximum value that can be measured with the resolution of our spectrometer (Figure \ref{fig:setup}(f)). However, in the remainder of this paper we will study cavities with typical $Q$ values below 1000, since these were more reliably fabricated in large numbers, permitting systematic studies. To transfer cavities, we press the GaP chip against a flexible polymer layer of Polydimethylsiloxane (PDMS), which separates the PC membranes from the chip while preserving their arrangement. The adhesion between the membranes and the PDMS is weak enough so that the GaP structures can be stamped onto the sample that is to be imaged, as shown in Figure \ref{fig:setup}(e). In our demonstration, the sample consists of  $\sim 30$ nm diamond nanocrystals that are dispersed on a glass slide covered by a $100$\unit{nm} thick layer of PMMA, for which the transfer process succeeds with $\sim 80\%$ probability for each membrane. The sample is mounted in a scanning confocal microscope with an oil immersion lens. A tungsten tip with radius $< 0.5$\unit{$\mu$m} is used to scan and position a PC nanocavity with nanometer resolution (Figure \ref{fig:setup}(g)).

\begin{figure}
%\centering{\includegraphics[width=.8\linewidth]{Fig1.pdf}}
\centering{\includegraphics[width=1\linewidth]{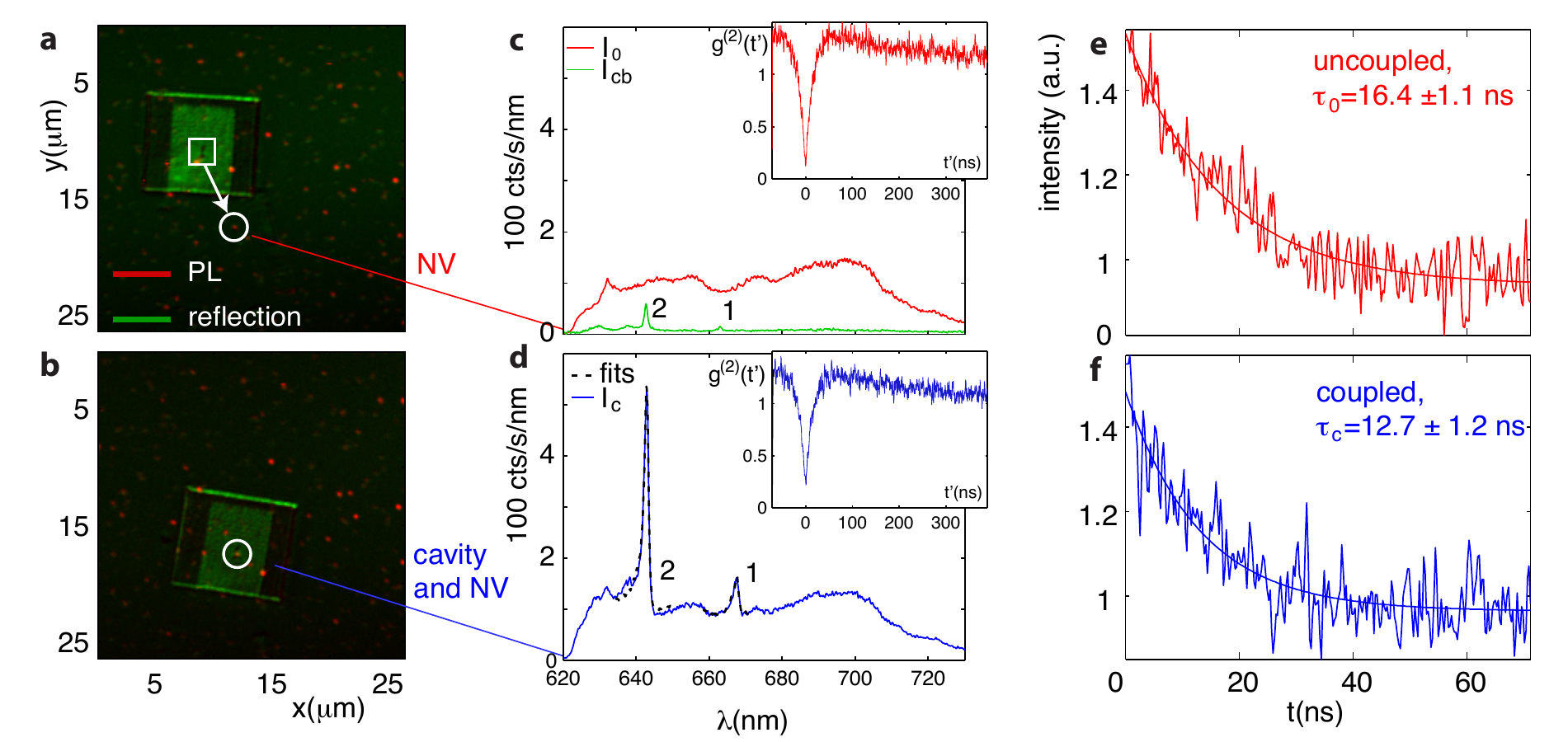}}
\caption{{\footnotesize The photonic crystal is moved from an initial uncoupled position (a) into alignment with the target NV center (b). The pump laser reflectivity is shown in green and the photoluminescence in red; pump laser power is $500$\unit{$\mu$W}, focused to $\sim 0.2$\unit{$\mu$m}. (c) PL spectrum of the uncoupled NV ($I_0$) and uncoupled cavity background ($I_{cb}$). A photon correlation measurement (see Appendix \ref{sec:correlation}) shows that the NV emission is strongly antibunched (inset); this feature is surrounded by photon bunching due to shelving in a metastable state of the NV emitter \cite{2001.PhysicaB.Wrachtrup.NV_fluorescence}. (d) PL spectrum $I_c$ of the coupled NV-cavity system, again strongly antibunched (inset).  A fit to theory (Eq.\ref{eq:S}) gives the SE rate into the cavity normalized by the background emission rate, $f^{c}(\lambda_2)=5.3, f^{c}(\lambda_1)=0.7$.  (e) Time-resolved emission for the uncoupled NV, far removed from the PC membrane, and (f) the coupled NV. The $6$\unit{ps} excitation pulse was generated by a frequency-doubled 1064\unit{nm} laser at 20 MHz repetition. }}\label{fig:2}. 
%\vskip2mm
\end{figure}

Figure \ref{fig:2}(a) shows a room-temperature photoluminescence (PL) image of a typical sample (red signal) obtained by scanning a green excitation laser across the surface with galvanometric mirrors. The PL spots in the image correspond to single or clusters of NV centers in diamond nanocrystals. The PC nanocavity can be located with the excitation laser reflection (green in Figure \ref{fig:2}(a)) as well as weak fluorescence originating from impurities in GaP and PMMA (green curve in Figure \ref{fig:2}(c)). The PC fluorescence clearly shows two resonances at $\lambda_1=667.3$\unit{nm} and $\lambda_2=643.0$ with quality factors $Q_{1}=550$ and $Q_2=610$. These quality factors are rather low because of variabilities in the fabrication; we saw no degradation due to the transfer onto the PMMA substrate. By collecting spectra at different points within the cavity,  we can identify peak $1$ as the fundamental mode depicted in Figure \ref{fig:setup}(c) and peak $2$ as two nearly degenerate, oppositely polarized higher-order modes of the cavity. The technique and results of the spatially resolved cavity spectrum are discussed in Appendix \ref{sec:cavity_modes}. 

%XXXX , which were also confirmed by reflectivity with a broad-band source\cite{2009.OpEx.Englund.X-switch-stark}

We achieve deterministic coupling between the NV and the nanocavity by first selecting a `target' NV (indicated in Figure \ref{fig:2}(a)). This center exhibits a broad spectrum $I_0$ (Figure \ref{fig:2}(c)), which is characteristic of NV centers and results from a broad phonon sideband extending from $\sim $ 640 to 800 nm. Importantly, the emission exhibits a strongly anti-bunched autocorrelation (inset), indicating that it results from a single emitter. To couple this emitter to the cavity, we position the PC membrane over the target NV using the tungsten tip. As shown in Figure \ref{fig:2}(d), the PL spectrum $I_{c}$ changes dramatically and shows strong peaks on resonance with the cavity modes. The intensities of these peaks are far higher than the cavity background, $I_{cb}$. Moreover, the autocorrelation of the coupled NV-cavity system is again strongly anti-bunched (Figure \ref{fig:2}(d)), indicating that it is driven by emission of the NV center. We also verified the electronic triplet state of the coupled NV by electron spin resonance measurements, as shown in Appendix \ref{sec:spin}.

The SE rate of an NV center is also modified by the presence of the PC slab. Specifically, Figures \ref{fig:2}(e) and (f) show that the lifetimes of the uncoupled and coupled NV centers are $\tau_{0,c}=16.4\pm  1.1, 12.7\pm 1.5$\unit{ns}, respectively. The PL spectra on and off the PC coupled with lifetime measurements allow the determination of the spectrally resolved SE rate enhancement, $F(\lambda)$, of the coupled emitter via the relation $F(\lambda) =  {I_{c}(\lambda) \tau_0}/{I_0(\lambda) \tau_{c}}$ (see Appendix \ref{sec:purcell}): the analysis of the data in Figure \ref{fig:2} yields $F(\lambda_1)=2.2$ and $F(\lambda_2)\sim 7.0$ (the full curve $F(\lambda)$ is plotted in the Appendix). 

We next demonstrate the spatial resolution of our method. By monitoring the fluorescence spectrum while scanning the cavity over the NV, we can map out the near-field emitter-cavity coupling. This is demonstrated in Fig 3(a-e), where we scan the cavity along its longitudinal (x-axis) over the sample in 3.4 nm steps. Figure 3(f) presents a series of PL spectra acquired as the cavity moves over the emitter, and reveals an intensity oscillation with a period corresponding to one PC lattice spacing, $a \sim 180$\unit{nm}. This oscillation corresponds to the spatially dependent SE modification, which is directly proportional to the cavity's electric field intensity.

\begin{figure}
%\centering{\includegraphics[width=.8\linewidth]{Fig1.pdf}}
\centering{\includegraphics[width=1\linewidth]{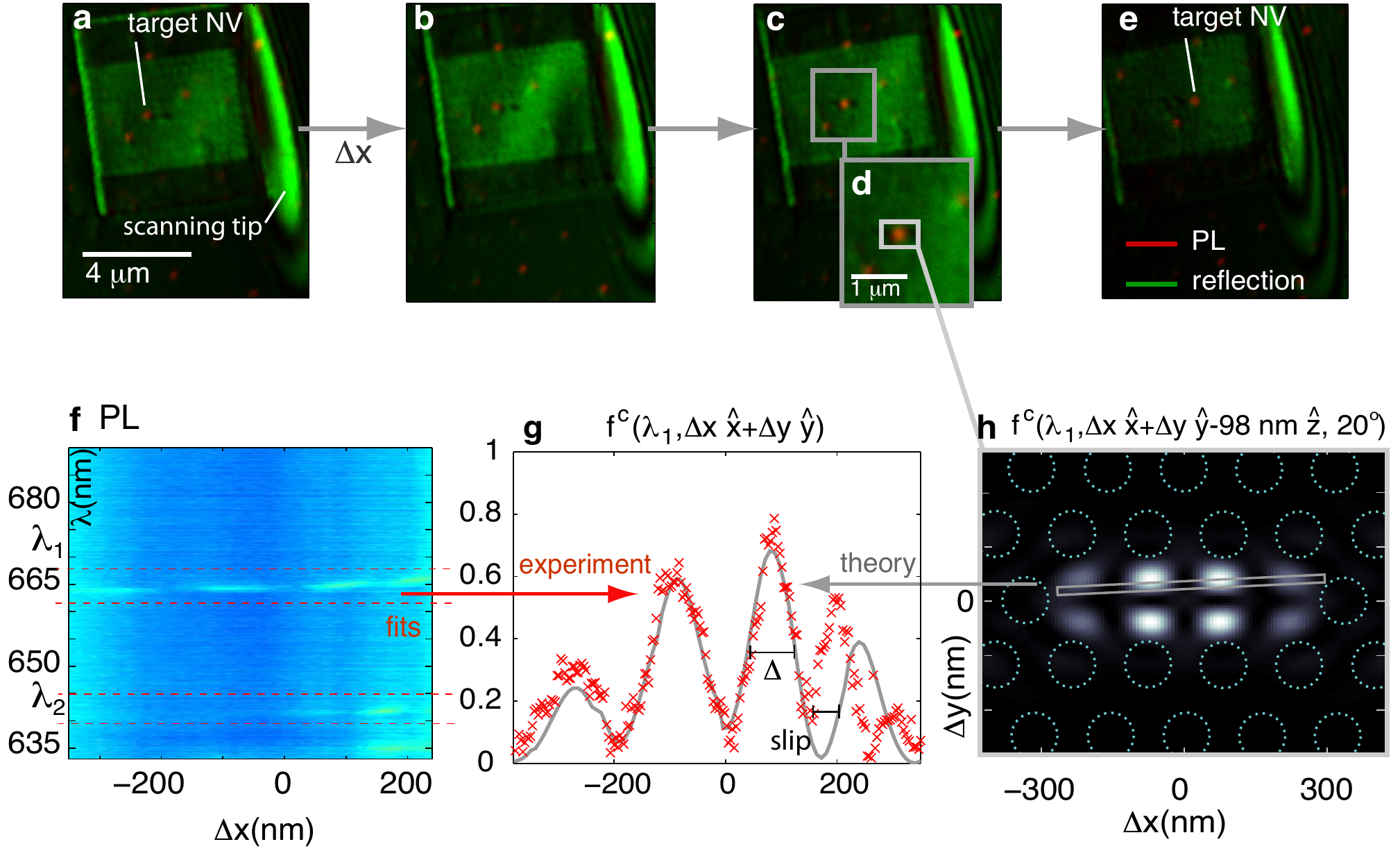}}
\caption{{\footnotesize Scanning of the PC nanocavity probe in small steps, shown in snap-shots (a-e).  (f) Photoluminescence scans for $\braket{\Delta x}= 3.4$\unit{nm} average step sizes.  (g) Fitted cavity SE rate enhancements $f^{c}(\lambda_1,\vec{r})$ for mode 1 showing a FWHM resolution of $\Delta\sim 80$\unit{nm}. (h) Expected SE enhancement factor $f^{c}(\lambda_1,\vec{r})$ and the estimated trajectory of the NV at $\Delta z= 98\pm 5$\unit{nm}, $\Delta y=70\pm 5$\unit{nm}, and $\vec{\mu}$ in the plane at 20$^{\circ}$ to the $x$-axis. The indicated track matches the observed SE enhancement in (g). }}\label{fig:3}
%\vskip2mm
\end{figure}

To analyze our observations, we note that the fluorescence of the coupled NV-cavity system is given by the emission directly from the NV, the emission through the cavity, and interference between the two: 
%. The contribution from each of these terms, $C_{NV}$, $C_{cav}$, and $C_{int}$, depends on the collection geometry. The expression for the spectrum detected on the spectrometer camera when the emitter is at position $\vec{r}$ is thus given by (derived in the Supplemental Information): 
\begin{equation}\label{eq:S}
S_d(\omega,\vec{r})= C_{NV}+C_{cav} f^{c}(\vec{r})|L(\omega)|^{2} + 2C_{int} \Re[ e^{i\Delta \phi}\sqrt{f^{c}(\vec{r})}L(\omega)],
\end{equation}
where $C_{NV}$, $C_{cav}$, and $C_{int}$ determine the relative contributions of the NV, the cavity, and their interference, respectively, which depend on the collection geometry and coupling to the collection fiber. This relation is derived in Appendix \ref{sec:spont_emission_mod}. $L(\omega)= 1/(1+i(\omega-\omega_c)/\kappa )$ gives the Lorentzian line shape of the cavity resonance at $\omega_c$ with linewidth $\kappa=\omega_c/2Q$, and $\Delta \phi$ accounts for the phase difference at the collection point between the direct NV emission and the emission through the cavity. The factor $ f^{c}(\omega,\vec{r}) $ is the SE rate enhancement of transitions in the phonon sideband of the NV with respect to the background emission rate into non-cavity modes.

% = \left|\vec{E}(\vec{r}) \cdot \vec{\mu}/(|\vec{E}_{\max}| |\vec{\mu}|) \right|^{2} f^{c,max} $ is the SE rate enhancement with respect to the background emission rate, where the first factor captures the spatial misalignment of the emitter, given by dipole $\mu$, from the cavity mode. $f^{c,max}$ is the maximum SE rate enhancement into the cavity mode, $F$, with respect to emission into non-cavity modes, $F_{PC,other}$: $f^{c,max}=F/F_{PC,other}$. %(We note that since the background emission rate can be different from the bulk emission rate in the vicinity of the PC slab, $f^{c}(\vec{r},\omega)$ is not identical to the SE modification with respect to the uncoupled emitter, $F(\omega)$, which was computed above.  They are related 

The coefficients $C_{NV}$, $C_{cav}$, and $C_{int}$ can be estimated from our experimental data as follows. Because of the high numerical aperture of our objective, nearly half of the emission from the cavity and the NV is collected: this observation suggests $C_{NV}\sim C_{cav}$. When the signal is collected through a single-mode fiber, the interference term represented by $C_{int}$ becomes important and results in Fano-like features in the spectrum (see Appendix \ref{sec:single_mode_scan}) \cite{2009.OpEx.Barclay-Santori-Painter}. However, we find that the interference term vanishes when a multi-mode fiber is used, and we can set $C_{int}=0$. A fit of Eq.\ref{eq:S} to the spectrum in Figure \ref{fig:2}(d) then yields $ f^{c}(\lambda=643\mbox{\unit{nm}},\vec{r})=5.3, f^{c}(667\mbox{\unit{nm}},\vec{r})=0.7$.

%XXXXThe signal $PL(\omega,\vec{r})$ in Fig.\ref{fig:3}(g) represents the convolution of the emitter's spatial profile, $e(\vec{r})$, and the spectrometer signal given in Eq.\ref{eq:S}. Thus, $PL(\omega,\vec{r}) =\int e(\vec{r}-\vec{r}') S_d(\omega,\vec{r}') dl'$, where $dl'$ is the displacement along the path of $\vec{r}'$. Since the NV center is less than 1 nm in size, we consider it as a delta function, so that $PL(\omega,\vec{r})=S_d(\omega,\vec{r})$, giving the response function of our scanning probe.  Once this response function is known, the SCN maps a general sample $e'(\vec{r})$ by first measuring the convolved image $PL'(\vec{r})$ and then deconvolving by the known response $S_d(\omega,\vec{r}')$ to obtain $e'(\vec{r})$. 

Since the signal in Figure \ref{fig:3}(f) is proportional to $S_d(\omega,\vec{r})$, we can now use Eq.\ref{eq:S} to compare the measured cavity signal to theory. Figure \ref{fig:3}(g) plots the fitted values of $ f^{c}(\omega_1,\vec{r})$ for the fundamental cavity mode frequency $\omega_1=2\pi c/\lambda_1$, as shown in the red crosses. By comparing the experimental  $ f^{c}(\omega_1,\vec{r})$ values to predictions for the cavity mode, we find a match between experiment and theory for an NV dipole $\mu$ that is $z=98\pm 5$\unit{nm} from the PC surface, as expected from the PMMA thickness, and at an angle of $20^{\circ}$ to the $x$-axis, obtained from the best fit to the data. For these conditions, the predicted value of SE rate modification corresponds to the track graphed in Figure \ref{fig:3}(h), in good agreement with experimental observations. A small discrepancy in the fit at $\Delta x\sim 190$\unit{nm} results primarily from positional slip of the PC cavity that can build up during the scan, a problem which could be improved by rigidly attaching the membrane to a stiffer scanning tip.% Note that the scan in Fig.\ref{fig:3}(g) characterizes the single-emitter response function of the SCN, which may be used to deduce the underlying image of complex samples (see Supplemental Information). %We note that Figs.\ref{fig:3}(g) and (h) represent the local density of states (LDOS) of our PC cavity probed by a single quantum emitter unlike previous LDOS measurements employing ensemble of emitters \cite{2005.Science.Noda.SE_control,Lodahl04,2005.PRL.Englund}.

The high spatial resolution and frequency-selective modification of spontaneous emission opens new possibilities for efficient interfacing of promising solid state qubits via optical fields. For instance, while the NV center is a promising system for quantum information processing, only the emission occurring into the zero phonon line (ZPL) is suitable for coherent optical manipulation. The frequency-selective emission enhancement demonstrated here potentially allows us to direct most of the emission of the selected NV centers into the ZPL. Furthermore, the hybrid approach is compatible with narrow linewidth NV emitters in bulk diamond at low temperature. This opens the door for applications ranging from quantum repeaters to single photon nonlinear optics. Moreover, although we have focused here on NV centers, our scanning technique provides a `cavity QED interface' that can be of use to a broad range of solid state qubits.

%Our demonstrated six-fold SE rate enhancement could be used to improve the decay branching ratios for spin manipulation and storage, while also allowing efficient collection from single spin systems. This is particularly important for pair-wise spin entanglement by photon scattering, whose success probability depends on the square of the collection efficiency. In the future, the SE enhancement can be improved by at least an order of magnitude through higher $Q$ and closer positioning to the cavity. The fraction of photons emitted into the cavity mode, $\beta$, is near $\sim 0.1$, and is estimated to be limited to $\sim 0.3$ because of the particularly broad homogeneous linewidth of the NV emitter.  The thermal broadening can be reduced below the cavity linewidth at low temperature in high-quality, bulk diamond. Our hybrid approach allows for deterministic coupling between bulk NV centres and a GaP PC cavity. For efficient coupling between individually addressable emitters on one chip, we are currently investigating the use of small photonic networks in the compact PC architecture. 
Furthermore, the PC scanning technique can serve as a new imaging approach with sub-wavelength resolution and high throughput, which we term a Scanning Cavity Microscope (SCM). Unlike other near-field probes that compromise the signal intensity to achieve high spatial resolution, SCM enables large count rates: in the demonstration shown here, we record up to $\sim 1\cdot 10^{6}$ photons/s from a single NV, exceeding collection rates reported to date with far field optics.  This can be further improved by efficiently out-coupling through cavity-coupled waveguides. In addition, the spatial resolution of the SCM is determined by the feature size of the confined field, which is $\Delta \sim 80$\unit{nm} for this cavity. This in-plane resolution may be improved substantially using cavity modes with small feature sizes, as in slot-waveguide cavities \cite{2005.PRL.Robinson-Lipson.slot-cav}. These qualities make the SCM a promising tool for label-free single molecule studies \cite{2007.Science.Armani-Vahala.sensing,2002.JPCB.Moerner.single-molecle-spectroscopy} or high-resolution studies of local index variations in thin films \cite{1979.ApplOpt.Pulker.optical_thin_films,2009.RevSciInstr.Bernal.SNOM_refr_index}. We note that a scan acquired by moving the photonic crystal membrane across the surface may need to be de-convolved by the response function of the cavity mode, as described in Appendix \ref{sec:general_samples}. Beyond high resolution and throughput, the SCM adds the capability to modify the spontaneous emission rate to near-field microscopy. This opens new possibilities for direct investigations of decay channels of optical emitters, such as light emitting diodes or fluorophores; for instance, by monitoring the emission intensity while effecting a known change in the radiative emission rate, the relative nonradiative recombation rate may be inferred, allowing a direct estimate of the radiative quantum efficiency of the material.

%\appendix{Appendix}
%\subsection{Lifetime measurements} 
%The pump pulses are generated with a Yb fiber laser at 1064 (Fianium), frequency doubled to 532\unit{nm}; pulse duration is 6\unit{ps}; repetition rate is down-sampled from 80\unit{MHz} to 20\unit{MHz} with an acousto-optic modulator.

\appendix 

\section{Wavelength-resolved excited state recombination rates} \label{sec:purcell}
%SHORTENED:

The excited state of the isolated NV$^{-}$ decays through a set of $N_{\lambda}$ radiative channels which we distinguish by their wavelength $\lambda_j$ and describe by the recombination rate $\gamma_j$, and nonradiative recombination described by a term $\gamma_{NR}$. We consider the NV at some position $\vec{r}_i$ from the cavity, and assume that it is pumped into the excited state at some rate $p_i$. Then the emission spectrum at wavelength $j$ is given by
\begin{eqnarray}\label{eq:I}
I_{ij}&=& c p_i \eta_{ij} \fr{\gamma_{ij}}{\sum_k \gamma_{ik}+\gamma_{NR}} \\
&=& c p_i \eta_{ij} \fr{\gamma_{ij}}{\Gamma_i}
\end{eqnarray}
where $\eta_{ij}$ is the collection efficiency, $\Gamma_i=\sum_k \gamma_{ik}+\gamma_{NR}$ is the excited state decay rate, and $c$ is a proportionality constant. The term $\beta_{ij}= \gamma_{ij}/\Gamma_i$ can be identified as the fraction of the total recombination rate that occurs through wavelength $j$. The decay rate can be obtained from a lifetime measurement, $\Gamma_i=1/\tau_i$. The ratio of the spectra at position $i$ and the `bulk position' $0$ (away from the PC slab) gives: 
\begin{eqnarray}\label{eq:F}
\fr{I_{ij}}{I_{0j}} &=&   \fr{p_i \eta_{ij} \gamma_{ij} \Gamma_0}{ p_0 \eta_{0j} \gamma_{0j}\Gamma_i} 
\end{eqnarray}
If the pump rate and collection efficiency are equal, we have
\begin{eqnarray}
\fr{I_{ij}}{I_{0j}} &=&   \fr{ \gamma_{ij} \Gamma_0}{\gamma_{0j}\Gamma_i} \\
&=&   F_{ij}  \fr{ \Gamma_0}{\Gamma_i},
\end{eqnarray}
where the Purcell factor $ F_{ij}=   \gamma_{ij}/ \gamma_{0j}=I_{ij}\Gamma_i/I_{0j}\Gamma_0 $.  For a position $i=1$ where the NV and cavity are aligned on top of each other, $F_{1j}=F(\lambda_j)$ is plotted in Figure \ref{fig:supp_purcell}. 

\begin{figure}
%\centering{\includegraphics[width=.8\linewidth]{Fig1.pdf}}
\centering{\includegraphics[width=1\linewidth]{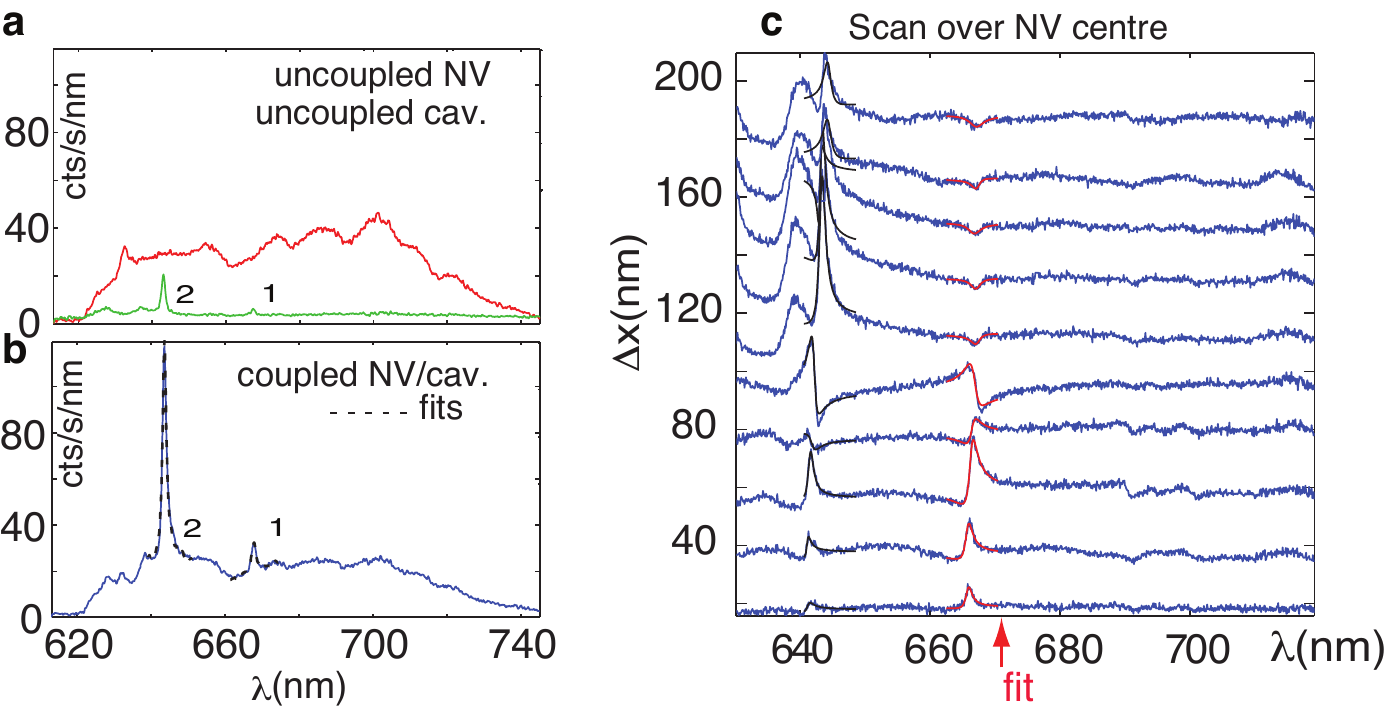}}
\caption{{\footnotesize Spectra collected into single mode fiber. (a) PL from uncoupled NV and cavity. (b) PL from cavity-coupled NV centre. (c) Scanning the cavity over the NV centre produces interference effects between NV and cavity emission. Fits to Eq.1 with $C_2=0.6$. }}\label{fig:supp_SMF_spectra}
%\vskip2mm
\end{figure}

%%Then we can rewrite Eq.\ref{eq:I} as 
%%\begin{eqnarray}
%%I_{ij}&=& c p_i \eta_{ij} \fr{F_{ij} \gamma_{i}}{\sum_k F_{ik} \gamma_{k}+\gamma_{NR}} \\
%%&=& p_i \eta_{ij}  \fr{F_{ij} A_j I_{0j} }{\sum_k F_{ik} A_j I_{0j}+\gamma_{NR}} ,\\
%%\end{eqnarray}
%%where we have used $\gamma_j = A_j I_{0j}$ with $A_j=\Gamma_0/p_0 \eta_{0j}$. Now solve for $\eta_{ij},...$ etc.. [[[to be finished]]]%Now, if the collection efficiencies $\eta_{ij}$ are equal or known, and $F_{ij}$ can be chosen such that there are $N_{\lambda}+2$ independent equations, the we can solve for $p_i \eta_{ij}, \gamma_{NR}$, and $A$, assuming $F_{ij}$ were previously obtained from the lifetime measurements described above. It is therefore possible with the scanning cavity to resolve the emission rates into radiative and nonradiative states. 

\begin{figure}
%\centering{\includegraphics[width=.8\linewidth]{Fig1.pdf}}
\centering{\includegraphics[width=.5\linewidth]{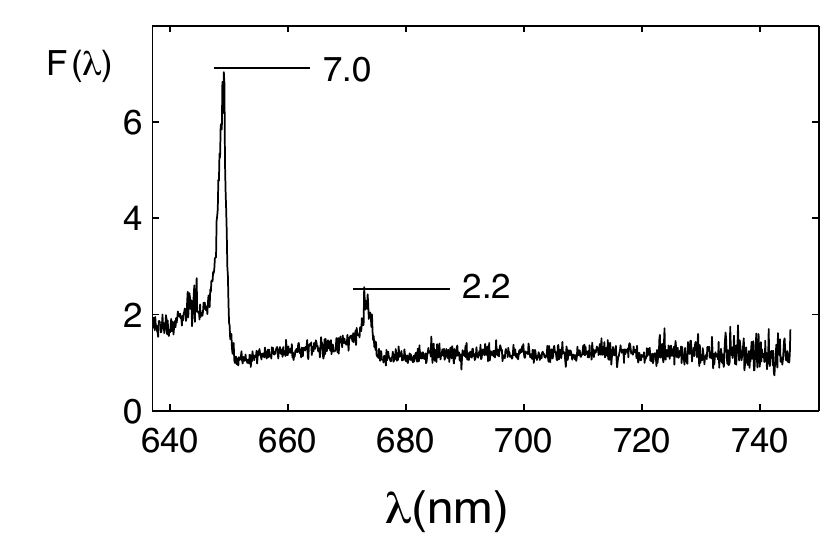}}
\caption{{\footnotesize The calculated spontaneous emission rate modification $F(\lambda)$.}}\label{fig:supp_purcell}
%\vskip2mm
\end{figure}

%\section*{Supplemental Material}
\section{Photon correlation measurements}\label{sec:correlation}
The photon correlation measurement is performed with a Hanbury Brown / Twiss  (HBT) setup. The PL is directed to a beam splitter, after which the two arms are coupled through single mode fibers (SMF) to single photon detector modules (Perkin Elmer). A time-correlated counting module is used to assemble the histogram of clicks on the two detectors. Figure \ref{fig:supp_SMF_spectra} shows the spectra collected through the SMF for the coupled and uncoupled positions of the cavity. 

\section{Characterization of cavity modes}\label{sec:cavity_modes}
\begin{figure}
%\centering{\includegraphics[width=.8\linewidth]{Fig1.pdf}}
\centering{\includegraphics[width=.9 \linewidth]{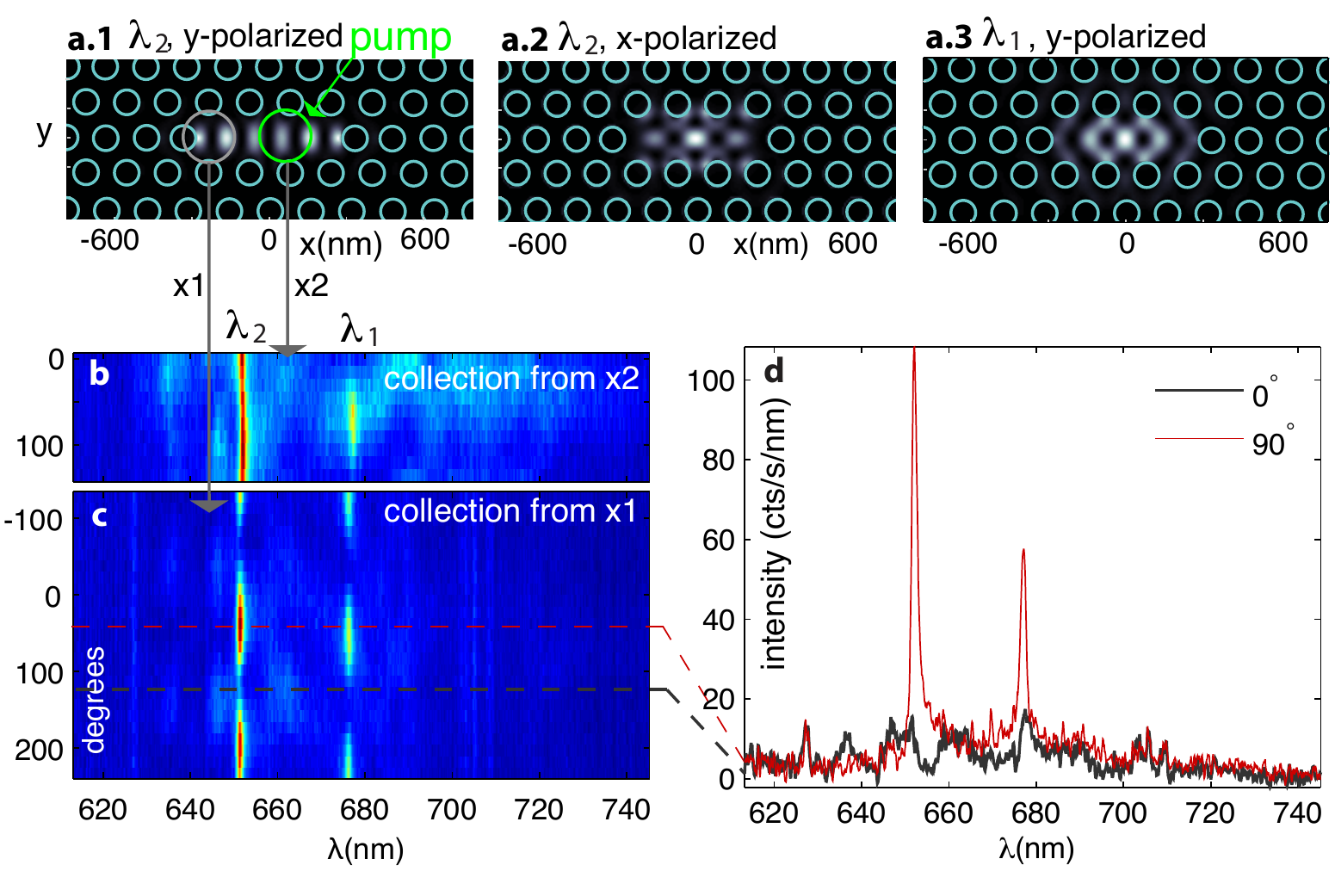}}
\caption{{\footnotesize Polarization dependence of cavity emission collected from spatially distinct points.  We position the cavity so that it slightly off-center from the NV emitter.  Exciting the NV and collecting from the point of excitation (a) we observe the fluorescence at 667nm to be strongly polarized while the fluorescence at 643\unit{nm} is only slightly polarized.  However, by collecting from the end of the cavity in (b-c), we can pick out the resonance at 643nm which is polarized along the same axis as the 667nm mode.}}\label{fig:4}
%\vskip2mm
\end{figure}

To further characterize the cavity resonances experimentally, we measured the polarization dependence of the emission collected from two different locations while pumping the NV near the center of the cavity. These positions are indicated as $x_1$ and $x_2$ in Figure \ref{fig:4}(a), together with the simulated electric field energy density for three relevant cavity modes. Collecting from the pump location $x_2$, we observe the mode at 667\unit{nm} to be strongly polarized, while the mode at 643\unit{nm} exhibits only weak polarization dependence (Figure \ref{fig:4}(b)), for a total of three modes. From $x_1$, we only observe polarized modes at $\lambda_1$ and $\lambda_2$. From these observations, we conclude that the peak at $\lambda_2$ corresponds to the two modes shown in Figure \ref{fig:4}(a.1,a.2) with nearly degenerate frequency (within the relatively broad linewidth), opposite polarization, and different spatial extent. The peak at $\lambda_1$ indeed corresponds to a single mode, which we identify as the lowest-frequency confined state of the L3 cavity.

% note: there is moduation on teh NV at a FSR ~ 14

\section{Single mode fiber spectra}\label{sec:single_mode_scan}
When the multi-mode fiber was replaced with a single mode fiber, the spectra in Fig. 2(c,d) in the main text show the same pattern, but at a reduced count rate, as displayed in Figure \ref{fig:supp_SMF_spectra}.

The simultaneous collection from the cavity and the emitter results in interference at the detector. As the cavity is scanned over the NV emitter, we observe Fano-like features at the cavity resonances shown in Figure \ref{fig:supp_SMF_spectra}(b), which are fit well by Eq.1  with $C_2=0.6$. Such interface effects have been described before in microdisks\cite{2009.OpEx.Barclay-Santori-Painter} and microspheres\cite{2008.NanoLett.Benson.microdisk_diamond_coupling} coupled to single NV emitters. In microscopy, the intereference signature may be used to better distinguish a single emitter from the background.  

\section{Electron Spin Resonance and Rabi Oscillations}\label{sec:spin}
\begin{figure}
%\centering{\includegraphics[width=.8\linewidth]{Fig1.pdf}}
\centering{\includegraphics[width=.9 \linewidth]{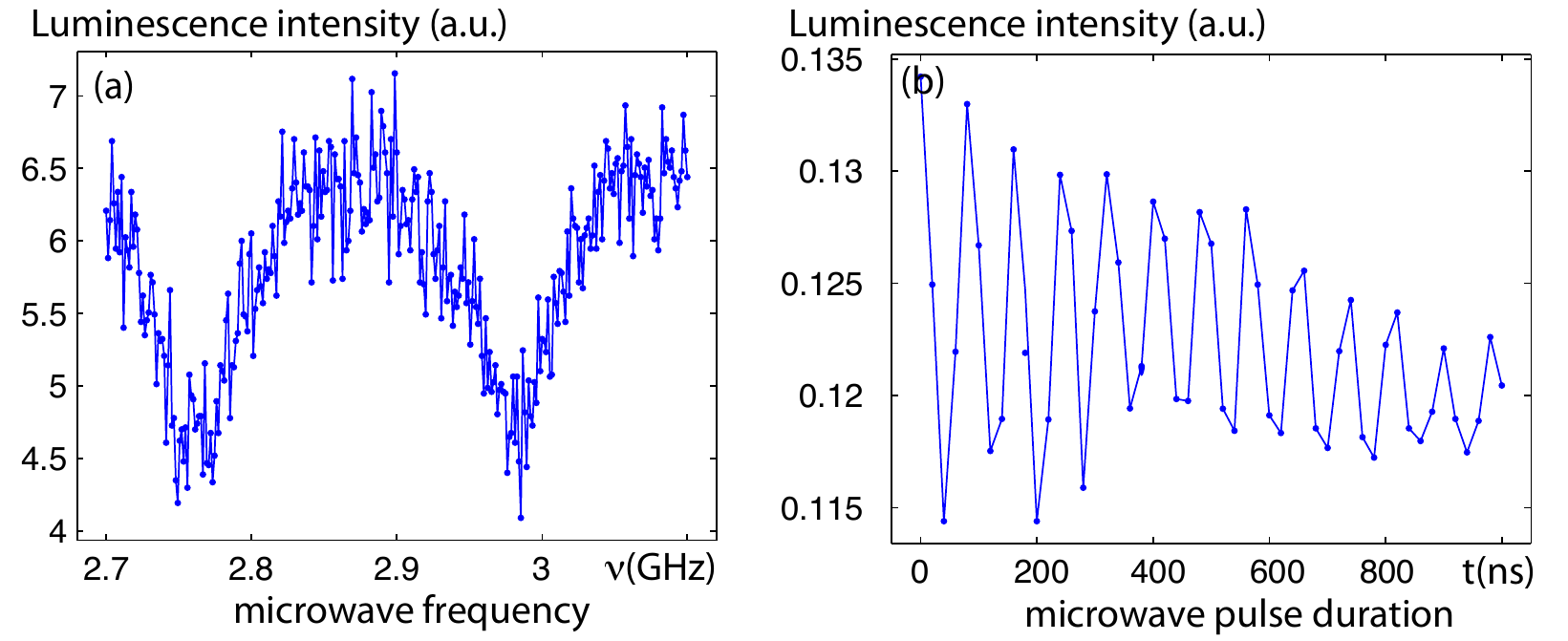}}
\caption{{\footnotesize (a) Spin transition observed in the NV fluorescence under excitation with a microwave field at frequency $\nu$. (b) Driven spin oscillations for $\nu=2.77$ GHz. }}\label{fig:ESR_NV}
%\vskip2mm
\end{figure}

The NV center has two unpaired electrons in a triplet configuration with a zero-field splitting of $\Delta=2.87$GHz between $m_s=0$ and $m_s=\pm 1$ sublevels. The excited state decay rates allow spin-sensitive detection since the fluorescence is reduced for the $m_s=\pm 1$ states \cite{2006.PRB.Manson.NV_model}. In the spin measurements described here, the $m_s=\pm 1$ sublevels were additionally split by an external magnetic field produced by a permanent magnet. Figure \ref{fig:ESR_NV}(a) shows the microwave transitions from $m_s=0$ to $m_s=\pm 1$, which are evident in the intensity of the cavity-coupled NV as a function of a microwave field of intensity $\nu$, while Figure \ref{fig:ESR_NV}(b) shows driven spin oscillations when the NV spin is optically initialized in $m_s=0$, excited with a microwave pulse for a duration $t$ at a frequency $\nu=2.77$ GHz, and optically detected using a pulse sequence described in elsewhere \cite{2006.Science.Childress.coherent_nuclear_spin_dynamics}.

\section{Spontaneous emission modification}\label{sec:spont_emission_mod}

%XXXXX*Supplemental materials: spontaneous emission modification section: I think your notation is a little confusing in that you define Purcell factor F, which I don't see in the expression, and don't define f_c

The Hamiltonian describing a NV (lowering operator $\sigma=\ket g \bra e$) coupled to a cavity mode (with annihilation operator $a$, without considering environment interactions, is given by
\begin{equation}
H=\frac{\Delta}{2}a^\dag a -\frac{\Delta}{2}(\sigma_z)+ig(\sigma a^\dag-a\sigma^\dag),
\end{equation}
where $\Delta$ is NV-cavity detuning and the reference energy is the mean of NV and cavity energy. 

The Master equation  is given by
\begin{equation}
\label{Maseq} 
\frac{d\rho}{dt}=-i[H,\rho]+\frac{\kappa}{2}(2a\rho a^\dag-a^\dag a \rho-\rho a^\dag a)+\frac{\gamma}{2}(2\sigma\rho \sigma^\dag-\sigma^\dag \sigma \rho-\rho \sigma^\dag \sigma)+\frac{\gamma_{d}}{2}(\sigma_z\rho\sigma_z-\rho)
\end{equation}
where $\gamma$, $\kappa$ and $\gamma_d$ account for NV population decay, cavity population decay and NV pure dephasing; and $\sigma_z=[\sigma^\dag,\sigma]$.

We consider the subspace of 0 or 1 excitation in the NV $\ket{g},\ket{e}$, and 0 or 1 excitations in the cavity $\ket{0},\ket{1}. $ From Eq. \ref{Maseq} we obtain the Maxwell Bloch equations\cite{2009.PRA.Poizat.dephasing}
\begin{eqnarray}
  \left \langle \frac{da}{dt} \right \rangle = (-i\frac{\Delta}{2}-\frac{\kappa}{2})\langle a \rangle+g\langle \sigma \rangle\\
  \left \langle \frac{d\sigma}{dt} \right \rangle = (i\frac{\Delta}{2}-\frac{\gamma}{2}-\gamma_d)\langle \sigma \rangle+g\langle \sigma_z a \rangle\\
\end{eqnarray}
These two coupled equations are solved with the initial condition $a(0)=0,\sigma(0)=1$; $\braket{\sigma_z a}=-a$ in the subspace considered here. We define
$c_1=\left(-i\frac{\Delta}{2}-\frac{\kappa}{2}\right)$ and
$c_2=\left(i\frac{\Delta}{2}-\frac{\gamma}{2}-\gamma_d\right)$ and
$\lambda_-=c_1+c_2-\sqrt{(c_1-c_2)^2-4g^2}$ ,
$\lambda_+=c_1+c_2+\sqrt{(c_1-c_2)^2-4g^2}$. 
The time-dependent solution is
\begin{equation}
a(t)=\frac{(e^{\lambda_+t}-e^{\lambda_-t})g}{\sqrt{(c_1-c_2)^2-4g^2}}
\end{equation}
and
\begin{equation}
\sigma(t)=\frac{(c_1-c_2)(e^{\lambda_-t}-e^{\lambda_+t})+\sqrt{(c_1-c_2)^2-4g^2}(e^{\lambda_-t}+e^{\lambda_+t})}{2\sqrt{(c_1-c_2)^2-4g^2}}
\end{equation}
The electric field emitted by the NV and cavity is described by 
\begin{equation}
E^{(+)}=\sqrt{\gamma}\sigma \hat{e}_{NV}+\sqrt{\kappa} a \hat{e}_{c} + c.c.,
\end{equation}
where $\hat{e}_{NV}$ and $\hat{e}_{c}$ describe the spatial profiles of the emission from the NV center and the cavity. The emission is collected through a lens into an optical fiber (either single mode or multi-mode) with a set of orthonormal modes \{$\hat{f}(\vec{k},\omega)$\}.  The field at the output of the fiber, which is directed at the detector, is given by 
\begin{equation}
E'^{+}=U_F ( \sqrt{\gamma}\sigma \hat{e}_{NV}+\sqrt{\kappa} a \hat{e}_{c}  ),
\end{equation}
where $U_F= \sum_{\vec{k},\omega} ( \cdot \hat{f}(\vec{k},\omega)) (\hat{f}(\vec{k},\omega)\cdot)$. The spectrum is obtained from the Quantum Regression theorem, $S(\omega)\propto \int_{-\infty}^{\infty} \int_0^{\infty} \braket{ E'^{+}(t) E'^{-}(t') }dt dt'$. We make the assumption that the linewidth of the NV center is much broader than the NV-cavity detuning and all other loss or coupling rates: $\omega=\omega_c,\kappa\ll \gamma_d,g\ll \gamma_d,\kappa$. After normalizing the spectrum by the bare NV emission spectrum, we obtain
\begin{eqnarray} \nonumber
S'(\omega)&\propto& \hat{e}_{NV}U_F\hat{e}_{NV}+2 \Re[ \hat{e}_{NV} U_F \hat{e}_{c} e^{i\Delta \phi}\sqrt{f^{c}(\vec{r})}\fr{1}{1+i(\omega-\omega_c)/\kappa } ]+\\
&&\hat{e}_{c}U_F\hat{e}_{c} f^{c}(\vec{r}) | \fr{1}{1+i(\omega-\omega_c)/\kappa} |^{2} ,
\end{eqnarray}
where $F=(g(\vec{r},\vec{\mu}))^{2}/\kappa\gamma$ denotes the Purcell factor at position $\vec{r}$ and orientation $\vec{\mu}/|\vec{\mu}|$ of the NV in the field of the cavity. In this experiment, it is not possible to calculate the term $\hat{e}_{NV} U_F  \hat{e}_{c}$ directly because of uncertainty in the coupling of the NV and cavity modes to the fiber. In general, we therefore represent the detected spectrum as 
\begin{eqnarray} \nonumber
S_d(\omega)= C_1+2C_2 \Re[ e^{i\Delta \phi}\sqrt{f^{c}(\vec{r})}\fr{1}{1+i(\omega-\omega_c)/\kappa } ]+C_3  f^{c}(\vec{r}) | \fr{1}{1+i(\omega-\omega_c)/\kappa} |^{2} ,
\end{eqnarray}
where the real coefficients $C_{i}$ are obtained by fits to the data. We obtain good fits for the single mode fiber with $C_2/C_1\sim 0.6$, and for the multi-mode fiber with $C_2/C_1\sim 0$. 

%So the spectrum emitted by the cavity $(S_{cav})$ and by the QD
%$(S_{QD})$ are given by (by virtue of the Quantum Regression theorem):
%\begin{equation}
%S_{cav}(\omega)=\frac{g^2}{(A-B)^2-4g^2}\left|\frac{1}{\omega-\lambda_-}-\frac{1}{\omega-\lambda_+}\right|^2
%\end{equation}
%\begin{equation}
%S_{QD}(\omega)=\frac{1}{2((A-B)^2-4g^2)}\left|\frac{A-B+\sqrt{(A-B)^2-4g^2}}{\omega-\lambda_-}-\frac{A-B-\sqrt{(A-B)^2-4g^2}}{\omega-\lambda_+}\right|^2
%\end{equation}

\section{Imaging of general samples}\label{sec:general_samples}
The signal $PL(\omega,\vec{r})$ in Fig. 3 in the main text represents the convolution of the emitter's spatial profile, $e(\vec{r})$, and the spectrometer signal given in Eq.1. Thus, $PL(\omega,\vec{r}) =\int e(\vec{r}-\vec{r}') S_d(\omega,\vec{r}') dl'$, where $dl'$ is the displacement along the path of $\vec{r}'$. Since the NV center is less than 1 nm in size, we consider it as a delta function, so that $PL(\omega,\vec{r})=S_d(\omega,\vec{r})$, giving the response function of our scanning probe.  Once this response function is known, the SCN maps a general sample $e'(\vec{r})$ by first measuring the convolved image $PL'(\vec{r})$ and then deconvolving by the known response $S_d(\omega,\vec{r}')$ to obtain $e'(\vec{r})$.

\end{document}